\documentclass[twocolumn, showpacs, preprintnumbers, superscriptaddress, aps, prl, floatfix, tightenlines]{revtex4-2}         
\usepackage{hyperref}
\usepackage{amsmath}
\usepackage{cleveref}
\usepackage{CJKutf8}
\usepackage{latexsym}
\usepackage{amstext}
\usepackage{amsfonts}
\usepackage{dsfont}
\usepackage{color}
\usepackage{amssymb}
\usepackage{relsize}
\usepackage{amsxtra}
\usepackage{verbatim}
\usepackage{hyperref}
\usepackage{graphics}
\usepackage{graphicx}
\usepackage{multirow}
\usepackage{ifpdf}

\renewcommand{\thetable}{\arabic{table}}

\begin{document}
\begin{CJK*}{UTF8}{shiai}

\title{Temperature-dependent structure and magnetization of YCrO$_3$ compound}

\author{Qian Zhao}
\thanks{These two authors contributed equally.}
\affiliation{Joint Key Laboratory of the Ministry of Education, Institute of Applied Physics and Materials Engineering, University of Macau, Avenida da Universidade, Taipa, Macao SAR 999078, China.}
\affiliation{State Key Laboratory of High Performance Ceramics and Superfine Microstructure, Shanghai Institute of Ceramics, Chinese Academy of Science, Shanghai 200050, China.}
\author{Yinghao Zhu}
\thanks{These two authors contributed equally.}
\affiliation{Joint Key Laboratory of the Ministry of Education, Institute of Applied Physics and Materials Engineering, University of Macau, Avenida da Universidade, Taipa, Macao SAR 999078, China.}
\affiliation{Guangdong--Hong Kong--Macao Joint Laboratory for Neutron Scattering Science and Technology, No. 1. Zhongziyuan Road, Dalang, DongGuan 523803, China.}
\author{Si Wu}
\affiliation{Joint Key Laboratory of the Ministry of Education, Institute of Applied Physics and Materials Engineering, University of Macau, Avenida da Universidade, Taipa, Macao SAR 999078, China.}
\affiliation{Guangdong--Hong Kong--Macao Joint Laboratory for Neutron Scattering Science and Technology, No. 1. Zhongziyuan Road, Dalang, DongGuan 523803, China.}
\author{Junchao Xia}
\affiliation{Joint Key Laboratory of the Ministry of Education, Institute of Applied Physics and Materials Engineering, University of Macau, Avenida da Universidade, Taipa, Macao SAR 999078, China.}
\affiliation{Guangdong--Hong Kong--Macao Joint Laboratory for Neutron Scattering Science and Technology, No. 1. Zhongziyuan Road, Dalang, DongGuan 523803, China.}
\author{Pengfei Zhou}
\affiliation{Joint Key Laboratory of the Ministry of Education, Institute of Applied Physics and Materials Engineering, University of Macau, Avenida da Universidade, Taipa, Macao SAR 999078, China.}
\author{Kaitong Sun}
\affiliation{Joint Key Laboratory of the Ministry of Education, Institute of Applied Physics and Materials Engineering, University of Macau, Avenida da Universidade, Taipa, Macao SAR 999078, China.}
\author{Hai-Feng Li}
\email{haifengli@um.edu.mo}
\affiliation{Joint Key Laboratory of the Ministry of Education, Institute of Applied Physics and Materials Engineering, University of Macau, Avenida da Universidade, Taipa, Macao SAR 999078, China.}

\date{\today}

\begin{abstract}

We have grown a YCrO$_3$ single crystal by the floating-zone method and studied its temperature-dependent crystalline structure and magnetization by X-ray powder diffraction and PPMS DynaCool measurements. All diffraction patterns were well indexed by an orthorhombic structure with space group of \emph{Pbnm} (No. 62). From 36 to 300 K, no structural phase transition occurs in the pulverized YCrO$_3$ single crystal. The antiferromagnetic phase transition temperature was determined as $T_\textrm{N} =$ 141.58(5) K by the magnetization versus temperature measurements. We found weak ferromagnetic behavior in the magnetic hysteresis loops below $T_\textrm{N}$. Especially, we demonstrated that the antiferromagnetism and weak ferromagnetism appear simutaniously upon cooling. The lattice parameters (\emph{a}, \emph{b}, \emph{c}, and \emph{V}) deviate downward from the Gr$\ddot{\textrm{u}}$neisen law, displaying an anisotropic magnetostriction effect. We extracted temperature variation of the local distortion parameter $\Delta$. Compared to the $\Delta$ value of Cr ions, Y, O1, and O2 ions show one order of magnitude larger $\Delta$ values indicative of much stronger local lattice distortions. Moreover, the calculated bond valence states of Y and O2 ions have obvious subduction charges.

\end{abstract}

\maketitle
\end{CJK*}



\section{I. Introduction}

It is well known that structure symmetry and crystallographic environment are decisive factors in determining physical properties of materials \textsuperscript{\cite{WangMicrostructure, he2012phase, zhang2019structural, wang2019isostructural}}. Therefore, solving the crystal structure of perovskite transition-metal oxides with general formula of ABO$_3$ plays a pivotal role in shedding light on the mechanism of intriguing properties. For ideal ABO$_3$ compounds, the cubic crystal structure (space group: \emph{Pm$\overline{3}$m}) is constructed by a 3D framework of corner-sharing octahedra with B cations accommodated at the center \textsuperscript{\cite{ardit2010elastic}}, which generates crystal field splitting of the fivefold-degenerate 3\emph{d} orbitals and splits the \emph{d}-level into twofold-degenerate $e_\texttt{g}$ and threefold-degenerate $t_{\texttt{2g}}$ levels. With the symmetry breaking in a tetragonal crystal environment, distorted BO$_6$ octahedra further lift the twofold $e_\texttt{g}$ ($d_{x^2 - y^2}$ and $d_{3z^2 - r^2}$) and the threefold $t_{\texttt{2g}}$ ($d_{xy}$, $d_{zx}$, and $d_{yz}$) degeneracies.

Chemical pressure imposed by coordination of different A and B cations or by doping would result in some degrees of mismatch between the equilibrium A-O and B-O bond lengths. Consequently, A ions will be inadequate for filling the space between BO$_6$ octahedra. Thus structure symmetry lowers. This kind of distortion can be quantitatively reflected by the value of tolerance factor $\emph{t}$ \textsuperscript{\cite{looby1954yttrium, kumar2008prediction, Li2008}}. An alternative to distorted BO$_6$ octahedra and lowered crystal structural symmetry is temperature variation. Therefore, temperature-dependent structural study is important for a full understanding of the novel physical properties such as multiferroics \textsuperscript{\cite{fiebig2002observation}}, dielectrics \textsuperscript{\cite{serrao2005biferroic}}, thermoelectricity \textsuperscript{\cite{Okuda2001}}, superconductivity \textsuperscript{\cite{Cava1988}}, and colossal magnetoresistance \textsuperscript{\cite{Li2006, Li2007_2}}.

Ferroelectric compounds are technologically interesting materials in data storage devices, catalysts, and sensors \textsuperscript{\cite{Scott1998, Vanaken2004, Zhao2014}}. YCrO$_3$ is one of the most significant compounds in perovskite-type oxides due to its potential coexistence of ferroelectricity and magnetism and its high thermal, electrical, and structural stability \textsuperscript{\cite{looby1954yttrium, Chakraborty2021, RN60}}. This compound shows a coexistence of antiferromagnetism and weak ferromagnetism and displays a spin-phonon coupling \textsuperscript{\cite{Sharma2014}}. A concept of ‘local non-centrosymmetry’ was proposed for the understanding of ferroelectricity observed in the centrosymmetric crystal structure of YCrO$_3$ \textsuperscript{\cite{Ramesha2007}}. This is different from conventional ferroelectric systems. The local electric polarization of YCrO$_3$ single crystals grown by the flux method is along the \emph{c} axis and the [110] direction in the $Pbnm$ symmetry \textsuperscript{\cite{Sanina2018}}, which depends on the direction of applied electric field with respect to the crystal axes and was attributed to two types of local polar domains: One is with structural distortions forming near Pb$^{2+}$ ions; the other is the magnetic phase separation domains forming near Pb$^{4+}$ ions. Therefore, the introduction of impurity Pb$^{2+}$ and Pb$^{4+}$ ions in the crystallographic position of Y$^{3+}$ sites results in the energetically beneficial processes of the formation of the local ferroelectric ordering in the isolated domains \textsuperscript{\cite{Sanina2018}}. Unfortunately, the study did not find any homogeneous ferroelectric orders in YCrO$_3$ single crystals from 5 to 350 K \textsuperscript{\cite{Sanina2018}}. So far, the crystal structure of YCrO$_3$ has been determined to be orthorhombic with space group of \emph{Pbnm} \textsuperscript{\cite{Geller1956, Sanina2018}}. It was reported that YCrO$_3$ compound displays a canted antiferromagnetic (AFM) structure with antisymmetric spin superexchanges, and the weak ferromagnetism is along the \emph{c} axis below $T_\textrm{N}$ \textsuperscript{\cite{Judin1966, Bertaut1966, Morishita1981, Ramesha2007}}. The ferroelectric relaxation was observed below $\sim$ 450 K \textsuperscript{\cite{serrao2005biferroic}}. However, the coupling mechanism of structure, ferroelectricity, and magnetism is still not clear. Previously, small single crystals of YCrO$_3$ were grown in millimeter size by the flux method \textsuperscript{\cite{Remeika1956, Grodkiewicz1966, Sanina2018, Todorov2011}}, however, impurities always exist.

\begin{figure}[!t]
\centering
\includegraphics[width = 0.48\textwidth] {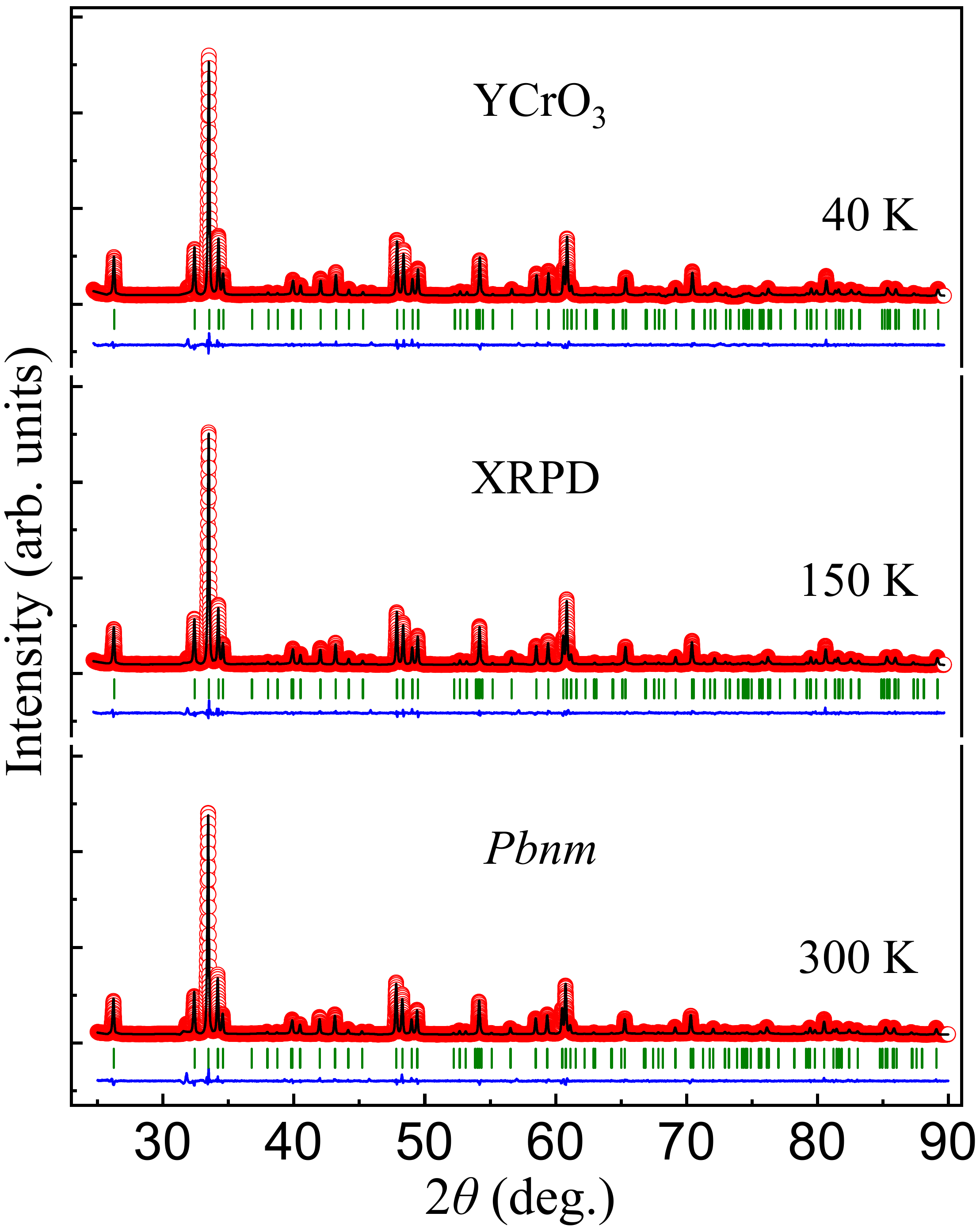}
\caption{
Observed (circles) and calculated (solid lines) XRPD patterns of a pulverized YCrO$_3$ single crystal, collected at 40, 150, and 300 K. The vertical bars mark the positions of Bragg reflections, and the lower curves represent the difference between observed and calculated patterns.
}
\label{XRPD}
\end{figure}

\begin{figure}[!t]
\centering
\includegraphics[width = 0.45\textwidth] {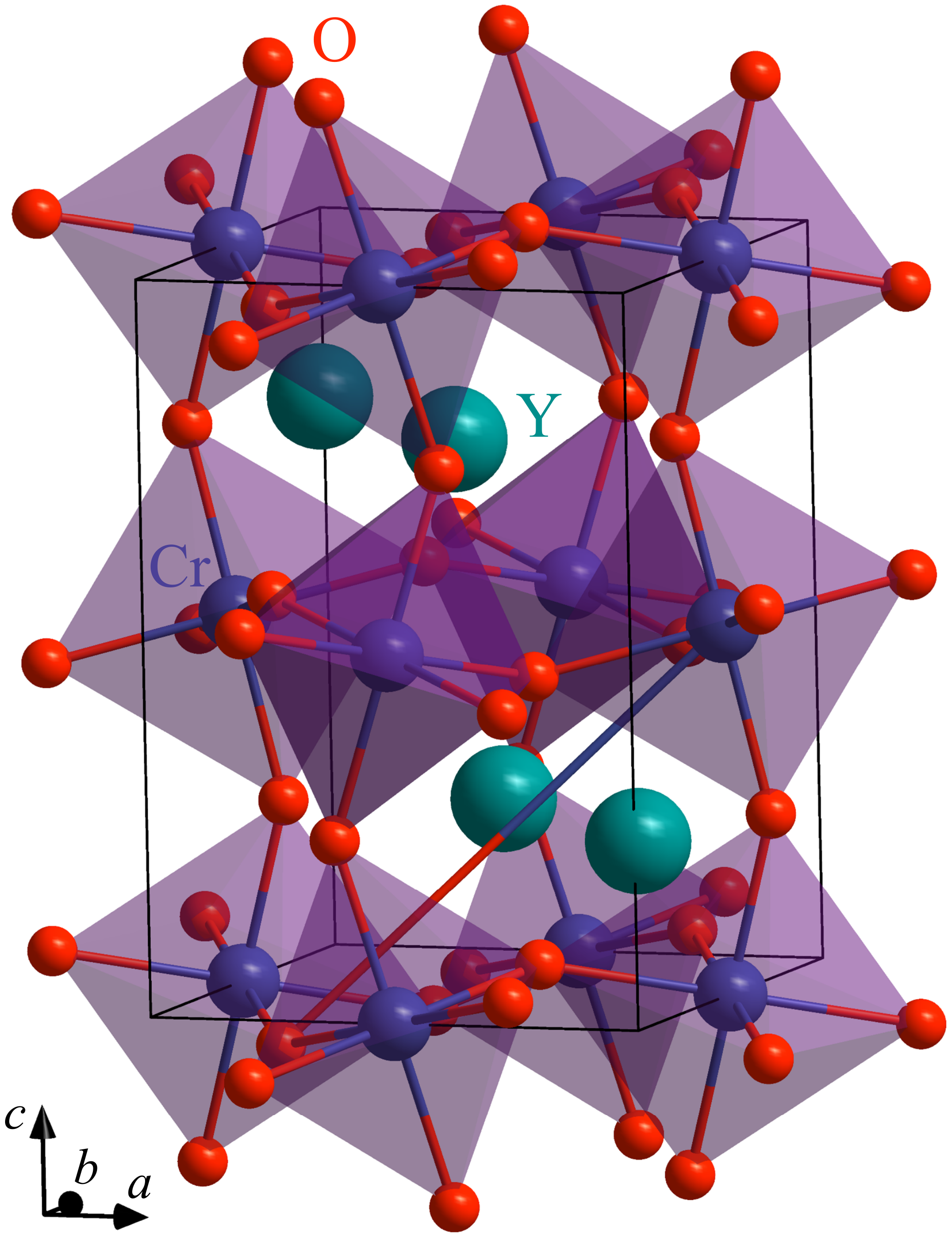}
\caption{
Orthorhombic crystal structure (space group: $Pbnm$) with one unit cell (solid lines) of a YCrO$_3$ single crystal at 300 K. The CrO$_6$ octahedra are highlighted in purple.
}
\label{unit-cell}
\end{figure}

\begin{table}[!t]
\renewcommand*{\thetable}{\Roman{table}}
\caption{Refined structural parameters of a pulverized YCrO$_3$ single crystal at 40, 150, and 300 K, including lattice constants, unit-cell volume, atomic positions, isotropic thermal parameter (\emph{B}), bond lengths, and the distortion parameter $\Delta$. We listed the Wyckoff site of each ion and the goodness of fit. The numbers in parenthesis are the estimated standard deviations of the (next) last significant digit.}
\label{RefinParas}
\begin{ruledtabular}
\begin{tabular} {llll}
\multicolumn {4}{c}{A pulverized YCrO$_3$ single crystal}                                                               \\
\multicolumn {4}{c}{(Orthorhombic, space group $Pbnm$ (No. 62), \emph{Z} = 4)}                                          \\ [2pt]
\hline
$T$ (K)                                       &40              &150              &300                                   \\ [2pt]
\hline
$a$ ({\AA})                                   &5.23006(6)      &5.23243(6)       &5.23942(7)                            \\
$b$ ({\AA})                                   &5.51712(7)      &5.51869(6)       &5.52065(7)                            \\
$c$ ({\AA})                                   &7.51809(9)      &7.52229(8)       &7.53183(9)                            \\
$\alpha (= \beta = \gamma)$ $(^\circ)$        &90              &90               &90                                    \\
$V$ ({\AA}$^3$)                               &216.934(5)      &217.215(4)       &217.858(5)                            \\ [2pt]
\hline
Y(4\emph{c}) \emph{x}                         &--0.01692(16)   &--0.01687(14)    &--0.01757(17)                         \\
Y(4\emph{c}) \emph{y}                         &0.06630(8)      &0.06707(8)       &0.06649(9)                            \\
Y(4\emph{c}) \emph{z}                         &0.25            &0.25             &0.25                                  \\
Y(4\emph{c}) \emph{B} (\AA $^2$)              &2.510(21)       &2.356(19)        &2.946(23)                             \\ [2pt]
\hline
Cr(4\emph{b}) $(x, y, z)$                     &(0.5, 0, 0)     &(0.5, 0, 0)      &(0.5, 0, 0)                           \\
Cr(4\emph{b}) \emph{B} (\AA $^2$)             &2.410(32)       &2.617(30)        &3.390(37)                             \\ [2pt]
\hline
O1(4\emph{c}) \emph{x}                        &0.11055(55)     &0.11940(52)      &0.10283(61)                           \\
O1(4\emph{c}) \emph{y}                        &0.45949(55)     &0.46754(52)      &0.46804(59)                           \\
O1(4\emph{c}) \emph{z}                        &0.25            &0.25             &0.25                                  \\
O1(4\emph{c}) \emph{B} (\AA $^2$)             &1.683(55)       &1.489(51)        &2.701(64)                             \\ [2pt]
\hline
O2(8\emph{d}) \emph{x}                        &--0.31168(43)   &--0.31187(40)    &--0.31179(47)                         \\
O2(8\emph{d}) \emph{y}                        &0.30031(45)     &0.30311(42)      &0.29542(52)                           \\
O2(8\emph{d}) \emph{z}                        &0.05282(31)     &0.05345(29)      &0.05465(35)                           \\
O2(8\emph{d}) \emph{B} (\AA $^2$)             &1.683(55)       &1.489(51)        &2.701(64)                             \\ [2pt]
\hline
Y-O11 (\AA)                                   &2.2694(33)      &2.3220(28)       &2.3046(34)                            \\
Y-O12 (\AA)                                   &2.2058(33)      &2.1511(27)       &2.2401(33)                            \\
Y-O21 (\AA) ($\times 2$)                      &2.4983(24)      &2.5034(22)       &2.4774(29)                            \\
Y-O22 (\AA) ($\times 2$)                      &2.2705(25)      &2.2610(23)       &2.2810(29)                            \\
$<$Y-O$>$ (\AA)                               &2.3355(11)      &2.3336(10)       &2.3436(12)                            \\ [2pt]
\hline
Cr-O1 (\AA) ($\times 2$)                      &1.9790(10)      &1.9897(9)        &1.9664(9)                             \\
Cr-O21 (\AA) ($\times 2$)                     &1.9678(26)      &1.9819(22)       &1.9498(27)                            \\
Cr-O22 (\AA) ($\times 2$)                     &2.0072(23)      &2.0014(21)       &2.0284(27)                            \\
$<$Cr-O$>$ (\AA)                              &1.9847(9)       &1.9910(7)        &1.9815(9)                             \\ [2pt]
\hline
$\Delta$(Y) $(\times 10^{-4})$                &25.264          &31.105           &16.950                                \\
$\Delta$(Cr) $(\times 10^{-4})$               &0.699           &0.163            &2.916                                 \\
$\Delta$(O1) $(\times 10^{-4})$               &38.737          &42.289           &53.244                                \\
$\Delta$(O2) $(\times 10^{-4})$               &96.382          &95.177           &91.470                                \\ [2pt]
\hline
$R_\textrm{p}$                                &3.14            &2.99             &3.05                                  \\
$R_\textrm{wp}$                               &4.57            &4.28             &4.68                                  \\
$R_\textrm{exp}$                              &2.88            &2.85             &2.90                                  \\
$\chi^2$                                      &2.53            &2.26             &2.61                                  \\
\end{tabular}
\end{ruledtabular}
\end{table}

In this paper, we have grown a centimeter-sized YCrO$_3$ single crystal with a laser-diode floating-zone (FZ) furnace \textsuperscript{\cite{Zhu2019}} and performed a temperature-dependent X-ray powder diffraction (XRPD) study from 36 to 300 K. We carried out magnetization measurements versus temperatures and applied magnetic fields with a PPMS DynaCool, with which we determined the onset temperature, $T_\textrm{N}$, of antiferromagnetism and observed weak ferromagnetism in the whole studied temperature regime below $T_\textrm{N}$. We refined the collected XRPD patterns and extracted crystallographic information such as lattice constants, unit-cell volume, local distortion parameter ($\Delta$), bond valence states (BVSs), and isotropic thermal parameter (\emph{B}). We found an anisotropic magnetostriction effect and subduction charges of Y and O2 ions.

\section{II. Experimental}

The polycrystalline YCrO$_3$ samples were synthesized by a mixture of raw materials Y$_2$O$_3$ (ALFA AESAR, 99.9\%) and Cr$_2$O$_3$ (ALFA AESAR, 99.6\%) compounds with a mole ratio of 1:1.08 through solid-state reaction method \textsuperscript{\cite{Li2008}}. Firstly, we mixed and ground the mixture of raw materials with a Vibrating Micro Mill (FRITSCH PULVERISETTE 0) for 1 h. Secondly, the mixture was heated in air at 1000$^{\circ}$C for 36 h with a temperature increasing/decreasing rate of 200$^{\circ}$C/h to perform the pre-reaction process. Thirdly, we performed a similar grinding, mixing, and heating procedure at 1200$^{\circ}$C for 36 h. After two cycles of calcination, we obtained a single-phase powder sample with green color. With additional treatments \textsuperscript{\cite{Zhu2019}}, the powder sample was isostatically pressed into a cylindrical rod of $\sim$ 12 cm long under a pressure of $\sim$ 70 MPa. The rod was then sintered in air at 1350$^{\circ}$C for 36 h. Through the above firing steps and grinding and mixing with a 50 mm diameter ball after each heating process, we finally prepared a dense and uniform sample \textsuperscript{\cite{Zhu2019-2}}. The single crystals of YCrO$_3$ compound were grown by the FZ method \textsuperscript{\cite{Li2008}} with a laser-diode FZ furnace (Model: LD-FZ-5-200W-VPO-PC-UM) \textsuperscript{\cite{Zhu2019, Wu2020}}.

We chose shiny pieces of YCrO$_3$ single crystals and carefully pulverized them into powder samples with a Vibratory Micro Mill and performed a XRPD study from 36 to 300 K with a 2$\theta$ range of 24--89$^\circ$ and a step size of 0.005$^{\circ}$ on an in-house X-ray diffractometer (Rigaku, SmartLab 9 kW) employing cooper $K_{\alpha1}$ (1.54056 {\AA}) and $K_{\alpha2}$ (1.54439 {\AA}) with a ratio of 2:1 as the radiation. XRPD patterns were collected at a voltage of 45 kV and a current of 200 mA at ambient conditions. The collected XRPD patterns were refined with the software of FULLPROF SUITE \textsuperscript{\cite{fullprof}}. We refined scale factor, lattice constants, background, peak profile shape, atomic positions, isotropic thermal parameter, and preferred orientation for following analyses. The magnetization measurements of a YCrO$_3$ single crystal were performed on a Quantum Design physical property measurement system (PPMS DynaCool instrument). The dc magnetization measurements at an applied magnetic field of 20 Oe were carried out with two modes from 1.8 to 350 K: One was after cooling down with zero magnetic field (ZFC), and the other was with the applied magnetic field while cooling (FC). Magnetic hysteresis loops were measured from 0 to 14 T, then to --14 T, and finally back to 14 T with temperatures at 5, 90, and 135 K.

\section{III. Results and discussion}

\subsection{A. X-ray powder diffraction}

In order to explore possible structural phase transitions in the pulverized YCrO$_3$ single crystal at low temperatures, we carried out XRPD measurements with increasing temperature from 36 to 300 K. The representative XRPD patterns as well as the corresponding structural refinements were shown in Fig.~\ref{XRPD}. The YCrO$_3$ compound undergoes an AFM phase transition \textsuperscript{\cite{Zhu2020}} at $T_\textrm{N} =$ 141.58(5) K. Below $T_\textrm{N}$, it enters into a canted AFM state. Therefore, we chose three temperature points 40 K (below $T_\textrm{N}$), 150 K (around $T_\textrm{N}$), and 300 K (above $T_\textrm{N}$) for the XRPD study. We indexed all collected data with an orthorhombic structure and a space group of \emph{Pbnm} (No. 62). We carefully checked the Bragg peak shape and positions, especially for the reflections locating at higher $2\theta$ degrees. No impurity peaks and detectable peak splitting were observed. This indicates that there is no structural phase transition occurring in the pulverized YCrO$_3$ single crystal in the whole studied temperature range within the present experimental resolution. The small difference between collected and calculated XRPD patterns (Fig.~\ref{XRPD}) and the low values of goodness of fit (Table~\ref{RefinParas}) further validate our refinements. The refined unit-cell of the orthorhombic structure was exhibited in Fig.~\ref{unit-cell}, and all extracted crystallographic information at 40, 150, and 300 K was listed in Table~\ref{RefinParas}.

\begin{figure}[!t]
\centering
\includegraphics[width = 0.48\textwidth] {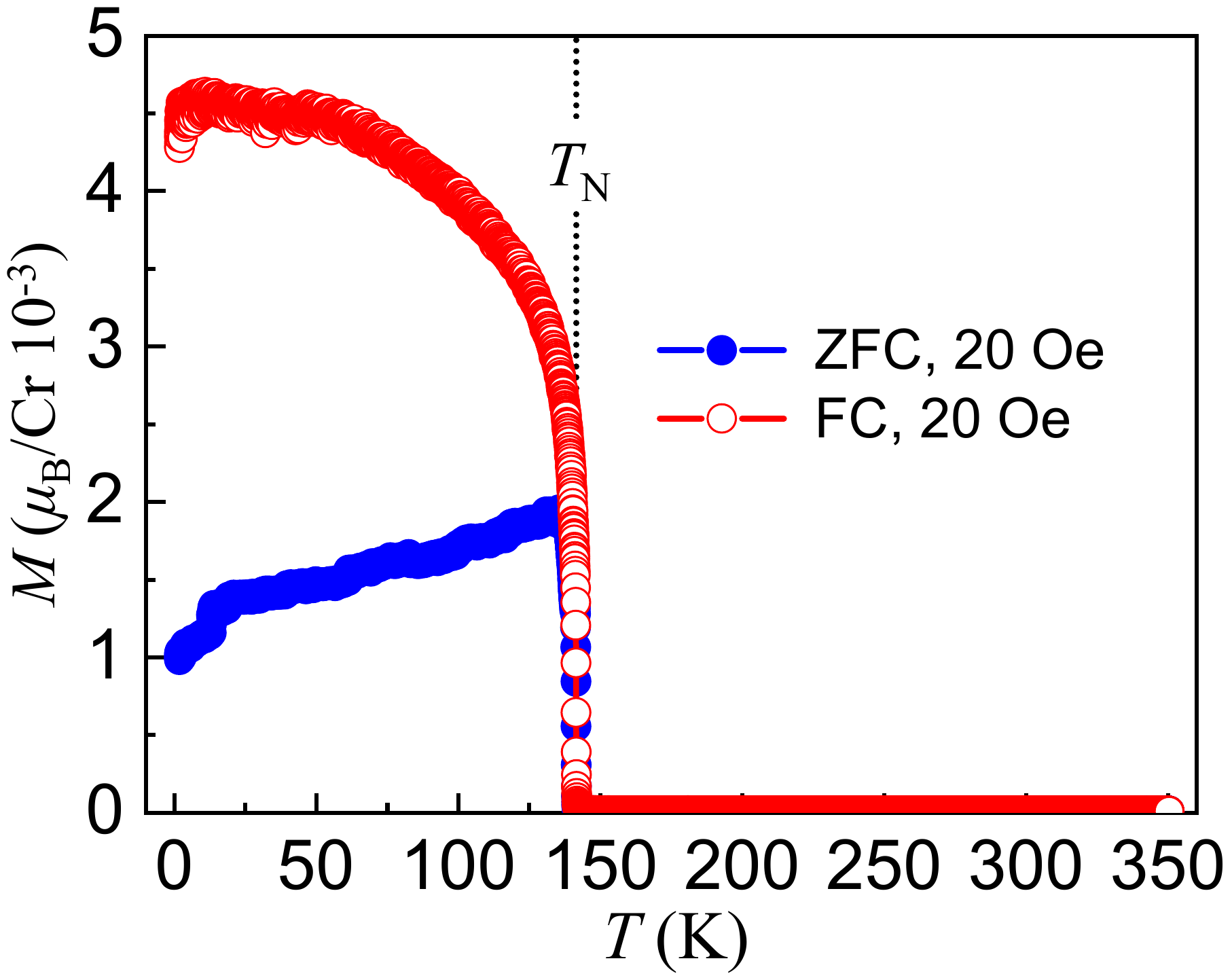}
\caption{
Temperature dependence of the ZFC and FC magnetization (\emph{M}) of chromium ions in a YCrO$_3$ single crystal measured at $B$ = 20 Oe.
}
\label{MT}
\end{figure}

\begin{figure}[!t]
\centering
\includegraphics[width = 0.48\textwidth] {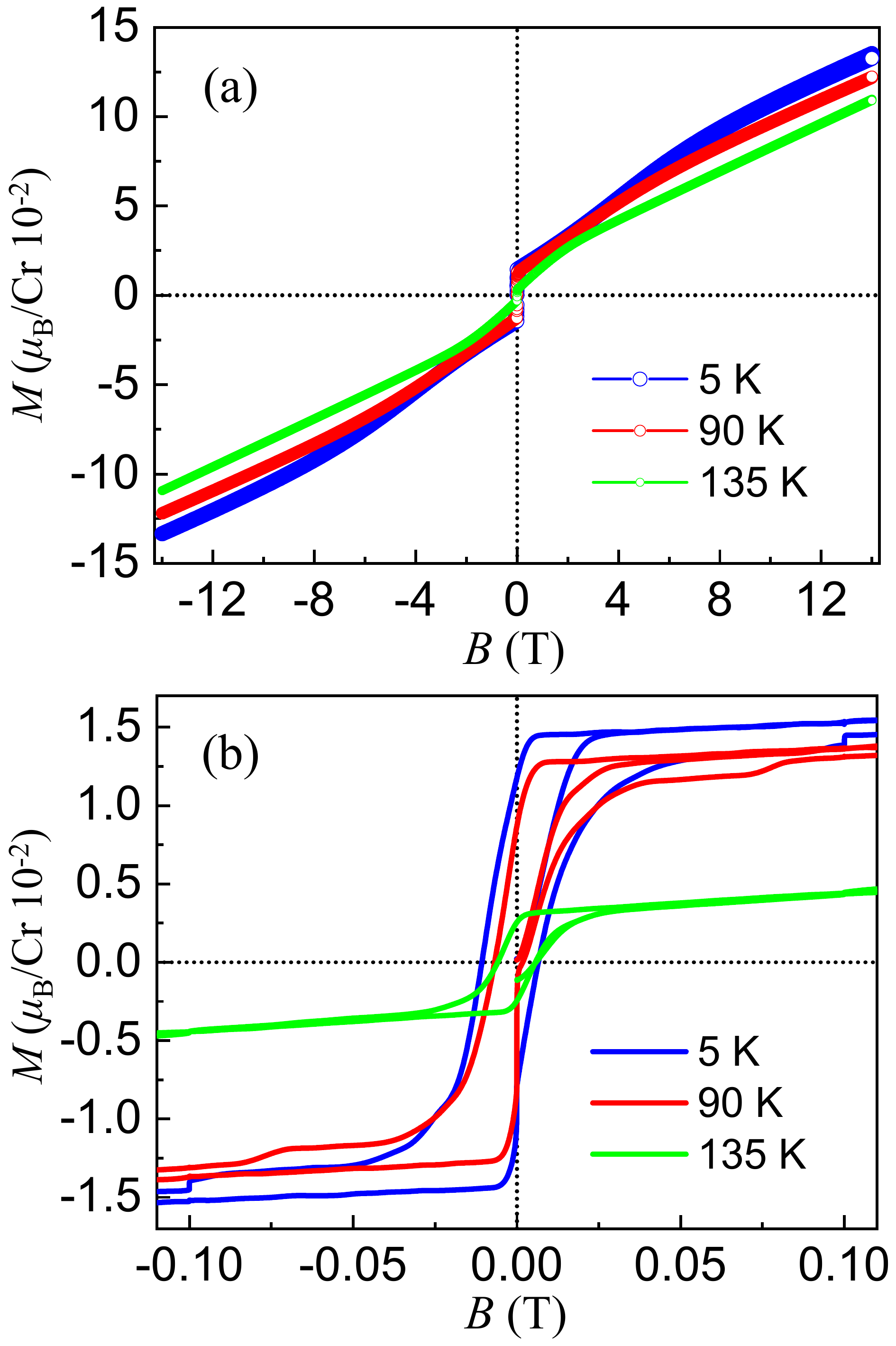}
\caption{
(a) Magnetic hysteresis loops of a YCrO$_3$ single crystal measured in a field range of --14 to 14 T at 5, 90, and 135 K. (b) For a clear comparison, we displayed the loops in a narrow field regime.
}
\label{MH}
\end{figure}

\begin{figure}[!t]
\centering
\includegraphics[width = 0.48\textwidth] {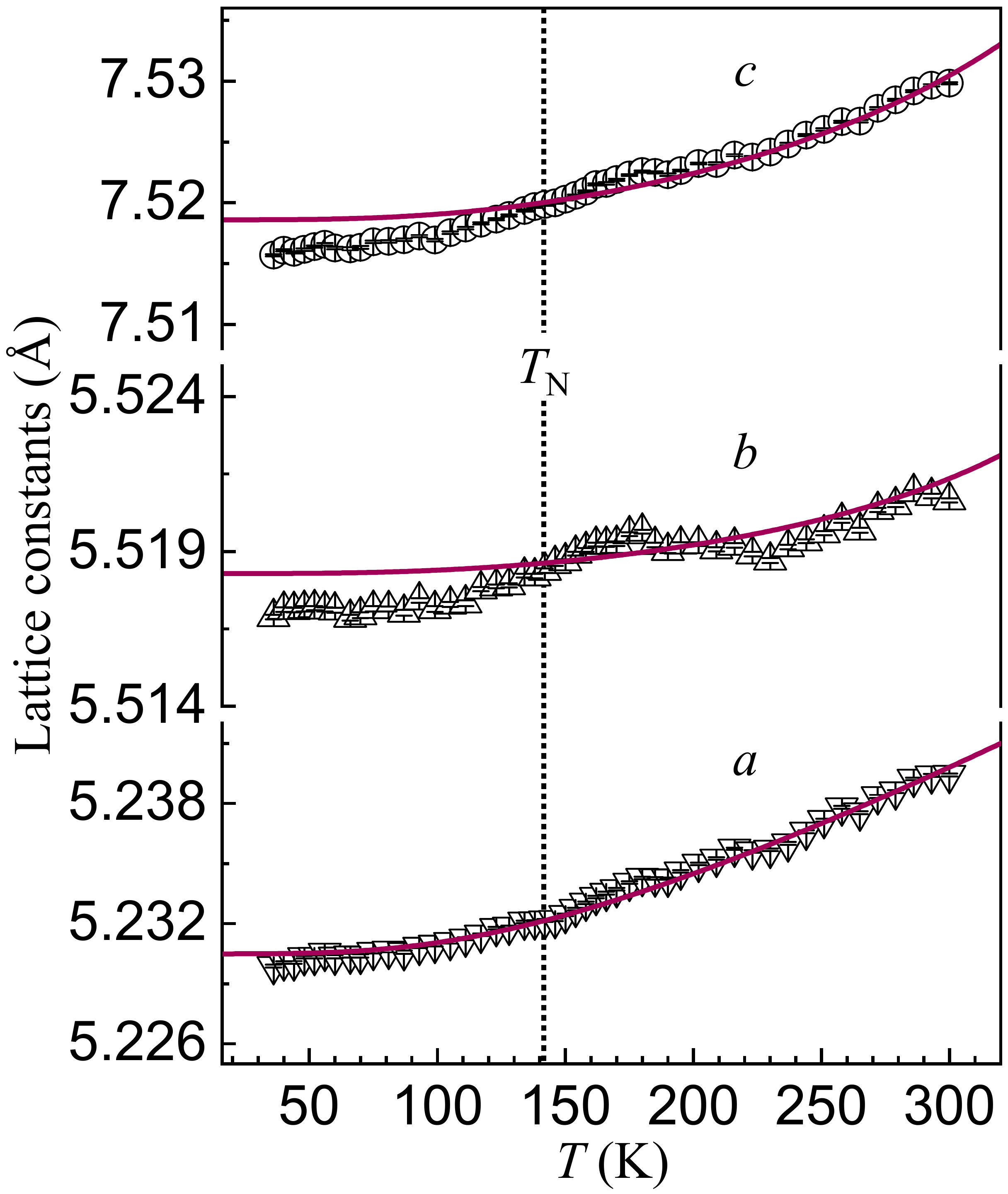}
\caption{
Temperature-dependent lattice constants \emph{a}, \emph{b}, and \emph{c} of a pulverized YCrO$_3$ single crystal (void symbols), which was extracted from our refinements based on the XRPD data collected from 36 to 300 K. The solid lines are theoretical estimates of the variation of structural parameters using the Gr$\ddot{\textrm{u}}$neisen model as discussed in the text with a Debye temperature $\theta_{D}$ = 580 K. The data was fit in paramagnetic state and extrapolated to overall temperature range. $T_\textrm{N}$ labels the AFM transition temperature. Error bars are the standard deviations obtained from the refinements.
}
\label{latticeP}
\end{figure}

\begin{figure}[!t]
\centering
\includegraphics[width = 0.48\textwidth] {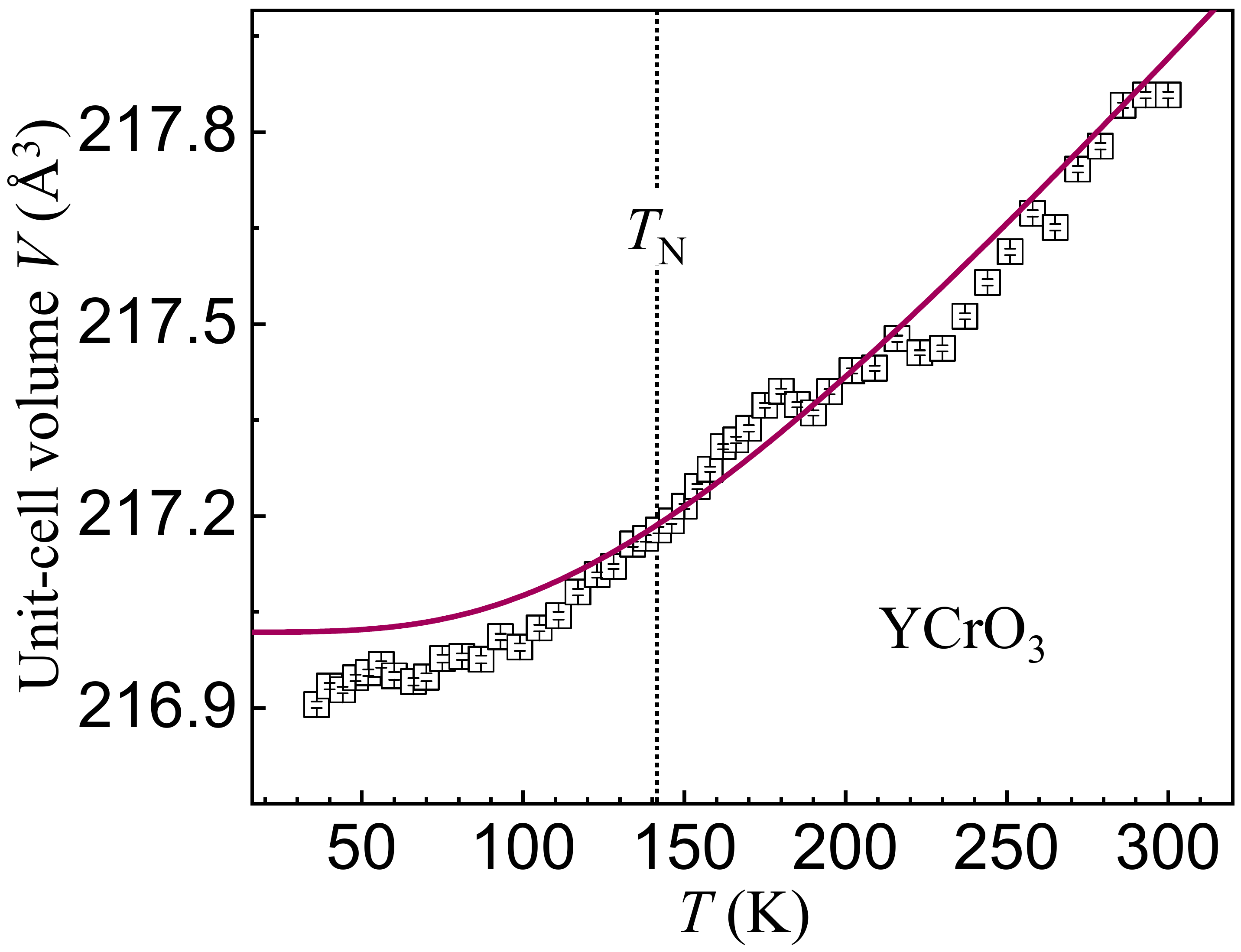}
\caption{
Temperature-dependent unit-cell volume, \emph{V}, extracted from the refinements. The solid line is theoretical estimate of the variation of unit-cell volume using the Gr$\ddot{\textrm{u}}$neisen model. We fit the data in paramagnetic state and extrapolated to overall temperature range. The vertical short-dotted line implies the AFM transition temperature $T_\textrm{N}$. Error bars are the standard deviations obtained from the refinements.
}
\label{UCVolume}
\end{figure}

\subsection{B. Magnetization}

Figure~\ref{MT} shows the magnetization versus temperature of a small YCrO$_3$ single crystal. We converted the unit of the vertical axis to $\mu_{\rm{B}}$ per Cr$^{3+}$ ion. The difference between ZFC and FC data is probably caused by the movement of magnetic domain walls instead of a spin–glass like transition \textsuperscript{\cite{chang2004magnetic}}. Upon cooling, the ZFC and FC magnetization curves remain extremely small until $\sim$ 141.58(5) K, around which the two curves increase simultaneously and sharply in a small temperature range of $\sim$ 5 K. With further cooling down to 1.8 K, a smooth increase appears in the FC curve, whereas an almost linear drop occurs in the ZFC curve. This indicates a reasonable canted AFM phase transition. At 1.8 K, ZFC magnetization = 1.02(1) $\times$ 10$^{-3}$ $\mu_{\rm{B}}$/Cr$^{3+}$.

Figure~\ref{MH}(a) shows the measurements of magnetic hysteresis loop at 5, 90, and 135 K. Figure~\ref{MH}(b) displays clearly the narrow quasi-parallelogram hysteresis loops in applied magnetic field range of $\sim$ --0.1 to 0.1 T. At 5 K, a coercive force of $\sim$ 84.26 Oe and a residual magnetization of $\sim$ 1.19(2) $\times$ 10$^{-2}$ $\mu_{\rm{B}}$/Cr$^{3+}$ were determined. An increase of temperature to 90 or 135 K diminishes the coercive force and the residual magnetization. The observed loops resemble the characteristic features of a soft ferromagnet at low temperatures. Thus there exists a coexistence of AFM and weak ferromagnetic behaviors in YCrO$_3$ below $T_\textrm{N}$ $\sim$ 141.58(5) K.

\subsection{C. Anisotropic magnetostriction effect}

The refined lattice parameters \emph{a}, \emph{b}, and \emph{c} from our XRPD studies were shown as void symbols in Fig.~\ref{latticeP}. The corresponding change in unit-cell volume \emph{V} was depicted in Fig.~\ref{UCVolume}. As is well known, the YCrO$_3$ compound is an insulator \textsuperscript{\cite{Zhu2020}}. Thus the contribution of electrons to the thermal expansion of lattice parameter ($\varepsilon$) can be neglected. The non-magnetic contribution to the thermal expansion is then mainly from phonons. Thus it can be approximately estimated according to the Gr$\ddot{\textrm{u}}$neisen rules at zero pressure with a second-order form \textsuperscript{\cite{Li2012, Wallace1998, Vocadlo2002}}, i.e.,
\setlength\arraycolsep{1.4pt} 
\begin{eqnarray}
\varepsilon(T) = \varepsilon_0 + \varepsilon_0\frac{U}{Q-BU},
\label{Gr1}
\end{eqnarray}
where $\varepsilon_0$ is the lattice parameter at 0 K, and the internal energy \emph{U} can be calculated with the Debye approximations \textsuperscript{\cite{Zhu2020}}
\begin{eqnarray}
U(T) = 9Nk_BT\left(\frac{T}{\Theta_D}\right)^3 \int^{\frac{\Theta_D}{T}}_0 \frac{x^3}{e^x - 1}dx,
\label{Gr2}
\end{eqnarray}
where \emph{N} = 5 is the number of atoms per formula unit, and $\Theta_D =$ 580 K is the Debye temperature \textsuperscript{\cite{Zhu2019-2}}. With equations~(\ref{Gr1}) and (\ref{Gr2}), we fit the lattice parameters of YCrO$_3$ in a temperature range of 150--300 K (in the paramagnetic state) and extrapolated the fits to overall temperatures shown as solid lines in Figs.~\ref{latticeP} and~\ref{UCVolume}. The values of the fitting parameters of $\varepsilon_0$, \emph{Q}, and \emph{B} were listed in Table~\ref{Gru}. It is obvious that below $T_\textrm{N}$, the fitting to the unit-cell volume deviates gradually from the refined values (Fig.~\ref{UCVolume}). Differences exist between refined values of lattice constants \emph{a}, \emph{b}, and \emph{c} and the corresponding theoretical estimates (Fig.~\ref{latticeP}), e.g., $\frac{a^\textrm{36K}_\textrm{Re} - a^\textrm{36K}_\textrm{Gr}}{a^\textrm{36K}_\textrm{Gr}}$ = --1.2(1) $\times$ 10$^{-4}$, $\frac{b^\textrm{36K}_\textrm{Re} - b^\textrm{36K}_\textrm{Gr}}{b^\textrm{36K}_\textrm{Gr}}$ = --2.6(1) $\times$ 10$^{-4}$, and $\frac{c^\textrm{36K}_\textrm{Re} - c^\textrm{36K}_\textrm{Gr}}{c^\textrm{36K}_\textrm{Gr}}$ = --3.9(1) $\times$ 10$^{-4}$. Therefore, a magnetostriction effect exists in YCrO$_3$, that is, the magnetic phase transition affects the thermal expansion of the lattice parameters. In addition, the effect is anisotropic. Below $T_\textrm{N}$, the magnetically-driven additional decreases of \emph{a}, \emph{b}, and \emph{c} jointly result in an extra contraction of the sample upon cooling, e.g., $\frac{V^\textrm{36K}_\textrm{Re} - V^\textrm{36K}_\textrm{Gr}}{V^\textrm{36K}_\textrm{Gr}}$ = --5.3(2) $\times$ 10$^{-4}$.

\begin{table}[!t]
\renewcommand*{\thetable}{\Roman{table}}
\caption{Values of the fitting parameters while estimating $a, b, c$, and $V$ of YCrO$_3$ compound with Eqs.~(\ref{Gr1}) and (\ref{Gr2}) in a temperature range of 150--300 K (Figs.~\ref{latticeP} and~\ref{UCVolume}). The numbers in parenthesis are the estimated standard deviations of the last significant digit.}
\label{Gru}
\begin{ruledtabular}
\begin{tabular} {llll}
Parameter                              &$\varepsilon_0$          &\emph{Q}                        &\emph{B}                                   \\
(unit)                                 &({\AA}/{\AA}$^3$)        &(J)                             &                                           \\ [2pt]
\hline
$a$                                    &5.23048(8)               &$1.62(1) \times 10^{-17}$       &13.42(6)                                   \\
$b$                                    &5.51828(6)               &$9.04(2) \times 10^{-17}$       &1407.68(7)                                 \\
$c$                                    &7.52059(8)               &$2.89(8) \times 10^{-17}$       &393.37(4)                                  \\
$V$                                    &217.018(2)               &$6.54(7) \times 10^{-18}$       &$-$9.79(7)                                 \\
\end{tabular}
\end{ruledtabular}
\end{table}

\subsection{D. Local distortions of Y, Cr, and O ions}

\begin{figure}[!t]
\centering
\includegraphics[width = 0.48\textwidth] {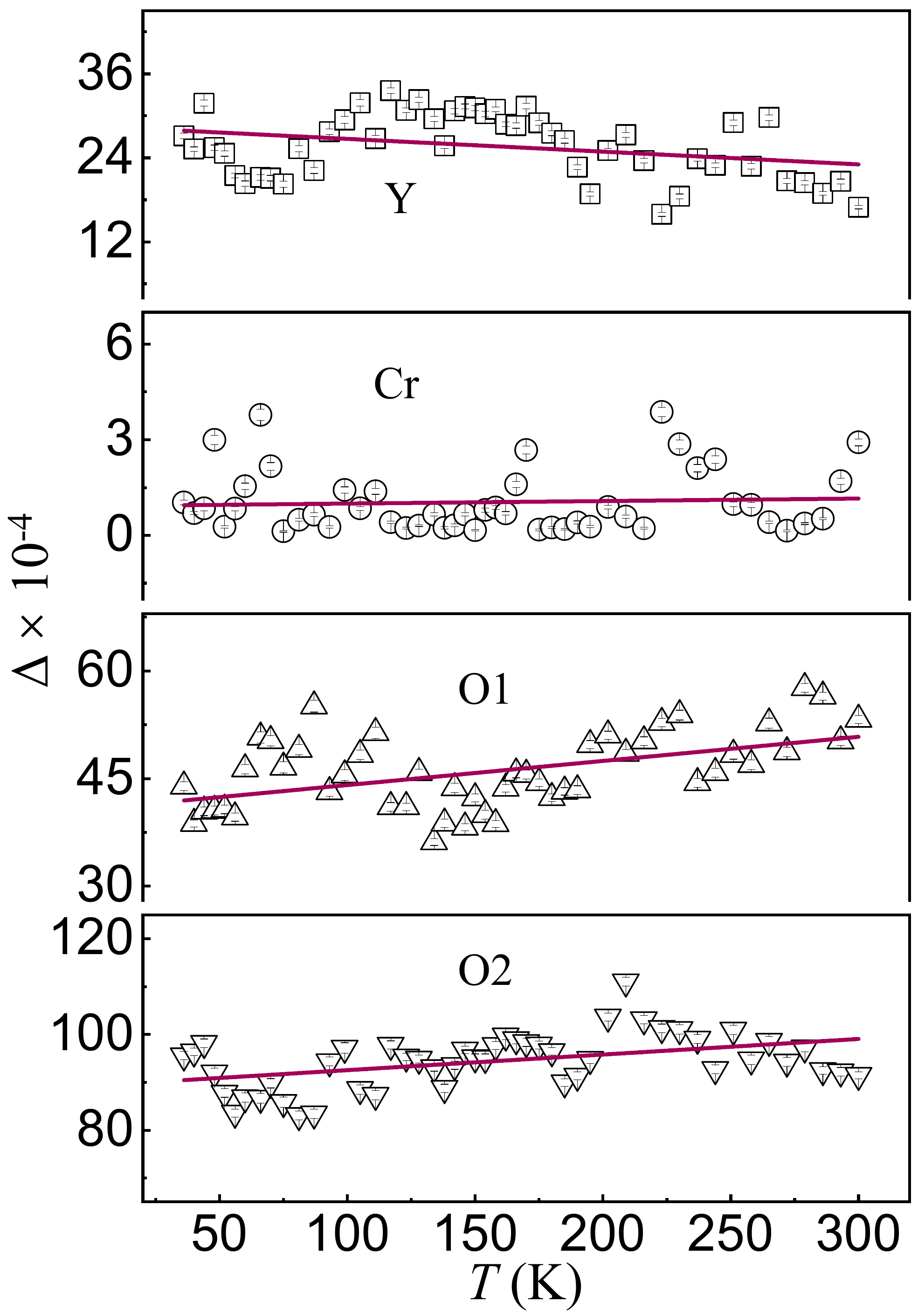}
\caption{
Temperature variation of the distortion parameter, $\Delta$, of Y, Cr, O1, and O2 ions of a single-crystal YCrO$_3$ (void symbols) from 36 and 300 K. Error bars are the standard deviations obtained from the refinements. The solid lines are tentative linear fits.
}
\label{distortion}
\end{figure}

Local crystalline environment can be quantitatively evaluated by the distortion parameter $\Delta$ \textsuperscript{\cite{Li2008, Li2009, Li2006, Li2007_2}}, i.e.,
\begin{eqnarray}
\Delta = \frac{1}{n}\sum_{i=1}^{n}\left(\frac{d_{i}-\langle d\rangle}{\langle d\rangle}\right)^2,
\label{delta}
\end{eqnarray}
where \emph{n} is the coordination number, \emph{d$_i$} is the bond length along one of the \emph{n} coordination directions, and $\langle d \rangle$ is the average bond length. We calculated the local distortion parameter $\Delta$ versus temperature for Y, Cr, O1, and O2 ions as shown in Fig.~\ref{distortion}. In principle, there is no Jahn-Teller effect for Cr$^{3+}$ ions. The calculated local distortion parameter $\Delta$ of Cr$^{3+}$ ions keeps almost a constant of $\sim$ 1.04329 $\times$ 10$^{-4}$ in the whole studied temperature range. By contrast, Y, O1, and O2 ions have much larger $\Delta$ values, more than one order of magnitude larger than that of Cr ions. For example, at 40 K, $\Delta$(O2) $\approx$ 2.48$\Delta$(O1) $\approx$ 3.81$\Delta$(Y) $\approx$ 137.88$\Delta$(Cr), indicating that there are significant local distortions of Y, O1, and O2 ions, which could break the local centrosymmetry and produce geometric ferroelectricity.

\begin{figure}[!t]
\centering
\includegraphics[width = 0.48\textwidth] {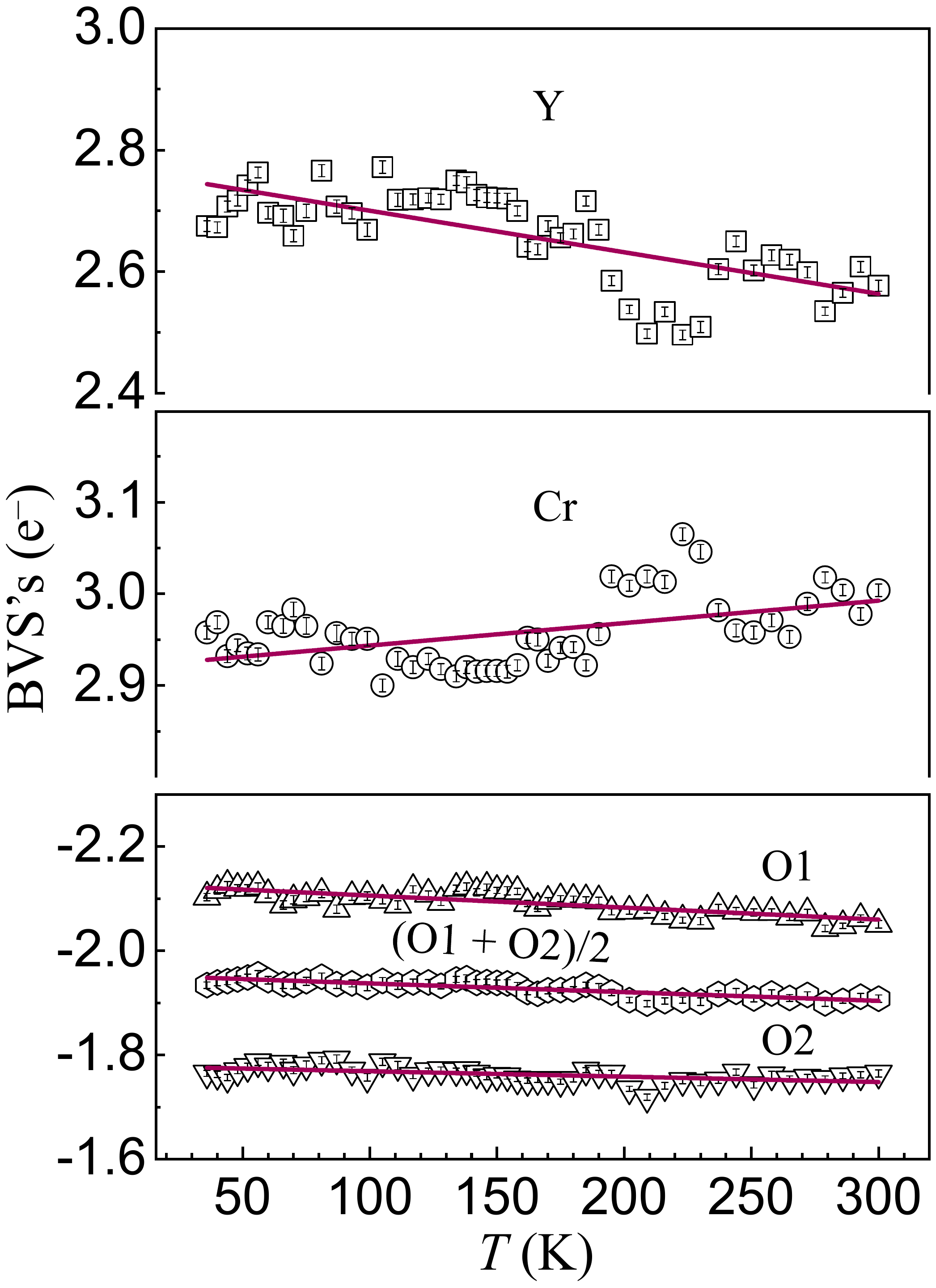}
\caption{
Temperature variation of the bond valence states of Y, Cr, O1, and O2 ions of a single-crystal YCrO$_3$ from 36 to 300 K (void symbols). We calculated the average bond valence states of O1 and O2 ions, i.e., (O1 + O2)/2. Error bars are the (propagated) standard deviations.
}
\label{BV}
\end{figure}

\begin{figure}[!t]
\centering
\includegraphics[width = 0.48\textwidth] {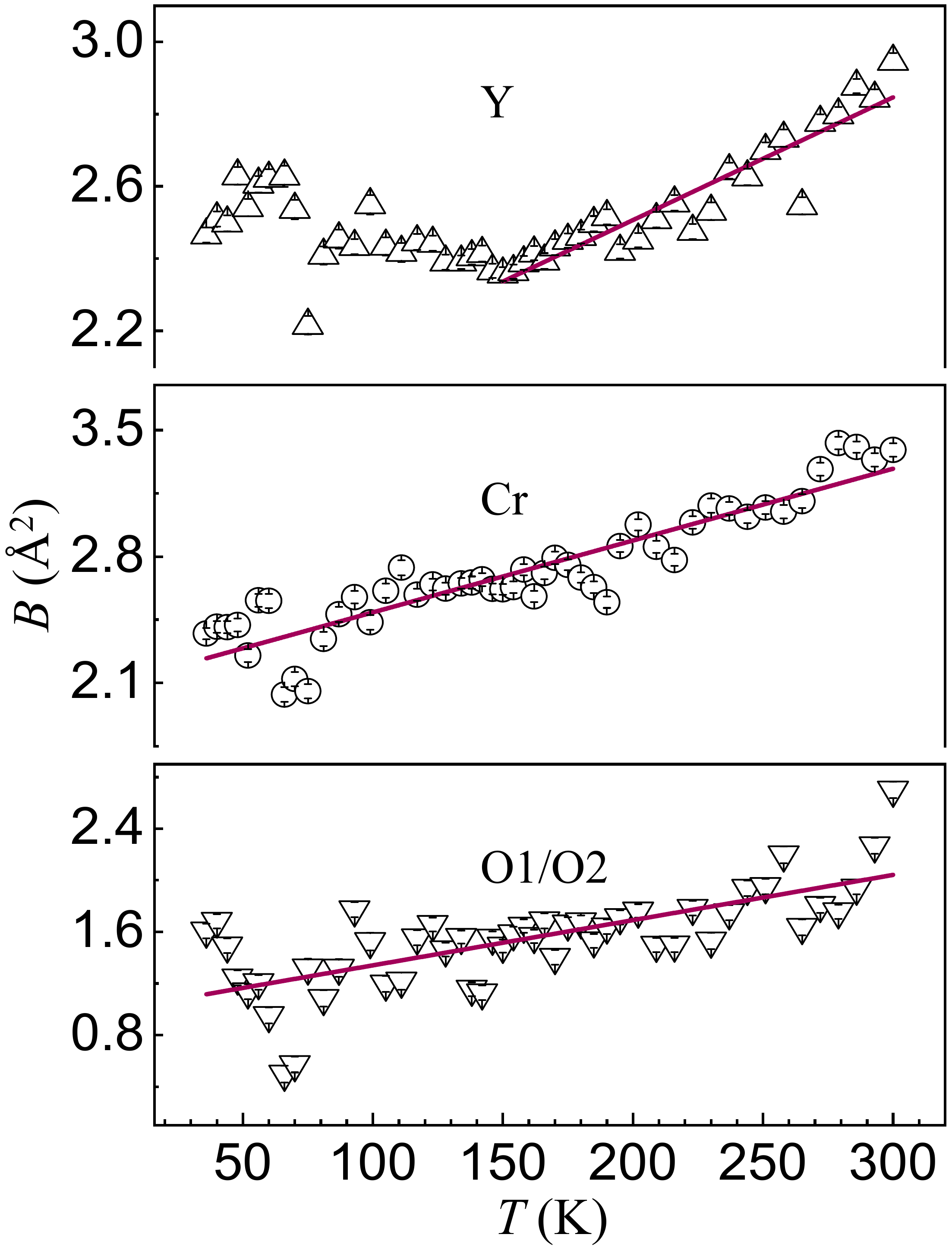}
\caption{
Temperature variation of the isotropic thermal parameters, \emph{B}, of Y, Cr, and O1/O2 ions of a single-crystal YCrO$_3$ (void symbols). During refinements, we constrained the \emph{B} values of O1 and O2 ions as the same. Error bars are standard deviations obtained from the refinements.
}
\label{thermalB}
\end{figure}

\subsection{E. Bond valence states of Y, Cr, and O ions}

BVSs are defined as the classical valence shared with each bond \textsuperscript{\cite{Mohri2000}}. From the refinements, we extracted the BVSs of Y, Cr, and O ions as shown in Fig.~\ref{BV}. The BVS of Y ions almost decreases linearly as temperature increases, whereas those of Cr and O ions nearly increase linearly. In the whole studied temperature range, the BVS of Y ions is largely smaller than $3+$, and that of O2 ions is much larger than $2-$. Therefore, it is clear that Y and O2 ions show subduction charges. We calculated the average BVS values of Cr and O1 ions, i.e., $2.958(7)+$ and $2.093(7)-$, respectively. By contrast, the average BVS values of Y and O2 ions are $2.661(9)+$ and $1.763(6)-$. They deviate largely from the ideal values of $3+$ (for Y ions) and $2-$ (for O2 ions). The two nonequivalent O1 and O2 ions show a charge difference $\sim$ 0.330 e$^-$ indicative of a big charge disproportion.

Figure~\ref{thermalB} shows the temperature-dependent values of isotropic thermal parameter \emph{B} of Y, Cr, and O1/O2 ions. During refinements, we constrained \emph{B}(O1) to be same as \emph{B}(O2). The parameter \emph{B} is called Debye-Waller factor or temperature factor, describing the attenuation of diffraction intensity caused by the thermal motion of regularly-arranged electrons in X-ray diffraction. As temperature increases, the \emph{B} values of Y (from 150 to 300 K), Cr, and O1/O2 ions (the whole temperature range) almost increase linearly (Fig.~\ref{thermalB}). This is consistent with the decrease in the X-ray diffraction intensities of Bragg peaks as shown in Fig.~\ref{XRPD}. It is noted that blow $\sim$ 150 K, the \emph{B} values of Y ions have no obvious changes.

It is stressed that the space group \emph{Pbnm} belongs to a centrosymmetric point group that has an inversion center. All building elements display inversion symmetry. Therefore, in such a space group, no net polar forms, thus no ferroelectricity appears. We observed much larger local distortions of Y, O1, and O2 ions and clear subduction charges of Y and O2 ions, therefore, we infer that there may exist local non-centrosymmetric domains. This could be the origin of the reported ferroelectricity observed in YCrO$_3$ compound.

\section{IV. Conclusions}

In summary, we carried out a detailed temperature-dependent XRPD study on a pulverized YCrO$_3$ single crystal grown by a laser-diode FZ technique. In the whole studied temperature range of 36--300 K, we did not find any crystalline structural phase transitions and thus indexed all collected XRPD patterns with an orthorhombic structure. By FULLPROF refinements, we extracted structure parameters of lattice constants, unit-cell volume, atomic positions, bond lengths, local distortions, BVSs, and thermal factors. We determined the AFM phase transition temperature $T_\textrm{N} =$ 141.58(5) K by the temperature-dependent magnetization study and found the weak ferromagnetism with the observation of magnetic hysteresis loops from 5 to 135 K. We observed an anisotropic magnetostriction effect below the AFM phase transition temperature as demonstrated by the deviations of lattice parameters (\emph{a}, \emph{b}, \emph{c}, and \emph{V}) from the Gr$\ddot{\textrm{u}}$neisen law. The local distortions ($\Delta$) of Y, O1, and O2 ions are much larger than that of Cr ions. We found that both Y and O2 ions display large charge displacements from theoretical values of $3+$ and $2-$, respectively. We suggest that local distortions of Y and O2 ions could be the reason for the observed ferroelectric characteristic in YCrO$_3$ compound.

\section{Competing interests}

The authors declare no conflict of interests.

\section{Acknowledgements}

Authors acknowledge the opening project of State Key Laboratory of High Performance Ceramics and Superfine Microstructure (Grant No. SKL201907SIC), Science and Technology Development Fund, Macao SAR (File Nos. 028/2017/A1 and 0051/2019/AFJ), Guangdong Basic and Applied Basic Research Foundation (Guangdong--Dongguan Joint Fund No. 2020B1515120025), and Guangdong--Hong Kong--Macao Joint Laboratory for Neutron Scattering Science and Technology (Grant No. 2019B121205003).

\bibliographystyle{iopart-num.bst}
\bibliography{YCOXRD}


\end{document}